\documentclass[12pt]{iopart}
\usepackage[utf8]{inputenc}
\usepackage{graphicx}
\usepackage{makecell}
\usepackage{url}

\begin{document}

\title{Accelerated iterative tomographic reconstruction with x-ray edge illumination}

\author{Peter Modregger}
\address{Department of Physics, Universit\"at Siegen, Walter-Flex-Straße 3, 57072 Siegen, Germany}
\address{FS-Petra, DESY Hamburg, Notkestraße 85, 22607 Hamburg, Germany} 
\ead{peter.modregger@uni-siegen.de}

\author{Tomasz Korzec}
\address{Department of Physics, Bergische Universit\"at Wuppertal,
Gau\ss strasse 20, 42119 Wuppertal, Germany}

\author{Jeff Meganck}
\address{Research \& Development, Discovery \& Analytical Solutions, PerkinElmer, Inc., 68 Elm Street, Hopkinton, MA 01748, United States of America}

\author{Lorenzo Massimi, Alessandro Olivo and Marco Endrizzi}
\address{Department of Medical Physics and Bioengineering, University College London, Gower Street, WC1E 6BT London, United Kingdom}

\vspace{10pt}
\begin{indented}
\item[]\today{} - version 2.0
\end{indented}

\begin{abstract}
Compared to standard tomographic reconstruction, iterative approaches offer the possibility to account for extraneous experimental influences, which allows for a suppression of related artifacts. However, the inclusion of corresponding parameters in the iterative forward model typically leads to longer computation times. Here, we demonstrate experimentally for phase sensitive X-ray imaging based on the edge illumination principle that inadequately sampled illumination curves result in ring artifacts in tomographic reconstructions. We take advantage of appropriately sampled illumination curves instead, which enables us to eliminate the corresponding parameter from the forward model and substantially increase computational speed. In addition, we demonstrate a 30\% improvement in spatial resolution of the iterative approach compared with the standard non-iterative single shot approach. Further, we report on several significant improvements in our numerical implementation of the iterative approach, which we make available online with this publication. Finally, we show that the combination of both experimental and algorithmic advancement lead to a total speed increase by one order of magnitude and an improved contrast to noise ratio in the reconstructions.
\end{abstract}

\vspace{2pc}

\section*{Introduction}

Differential phase contrast X-ray imaging techniques exploit phase in addition to absorption information about the sample. Especially for soft tissue samples, phase contrast can be considerably stronger than absorption contrast, which led to increasing interest in these techniques from the biomedical research community~\cite{Bravin2013a}. Among differential phase contrast X-ray techniques at least three different methods can be distinguished: analyser-based imaging (ABI)~\cite{Chapman1997,Oliva2020}, grating inteferometry (GI)~\cite{David2002,Momose2003a,Birnbacher2016} and edge-illumination (EI)~\cite{Olivo2007b,Olivo2021}.

The combination of differential phase contrast imaging with tomography poses a particular challenge as the explicit or implicit integration, which is necessary for differential phase methods, typically leads to strong artifacts~\cite{Thuring2015a}. This challenge was addressed in several ways, which included non-iterative~\cite{Pfeiffer2007c,Massimi2020,Massimi2021} and iterative integration~\cite{Thuring2015a,Sperl2014,Nilchian2015a} of retrieved differential projections and subsequent tomographic reconstruction. Alternatively, iterative tomographic reconstruction of either retrieved differential signals~\cite{Kohler2011b,Xu2012a,Nilchian2013b,Gaass2013a,Fu2013a} or based on various forward models~\cite{Ritter2013,Hahn2015,Brendel2016,Chen2017a,Teuffenbach2017a} were investigated. The latter are of particular interest as they offer the potential for utilizing fewer viewing angles and, thus, decreasing scan times and delivered dose.

\begin{figure}[htbp]
\begin{center}
\includegraphics[width=0.7\textwidth]{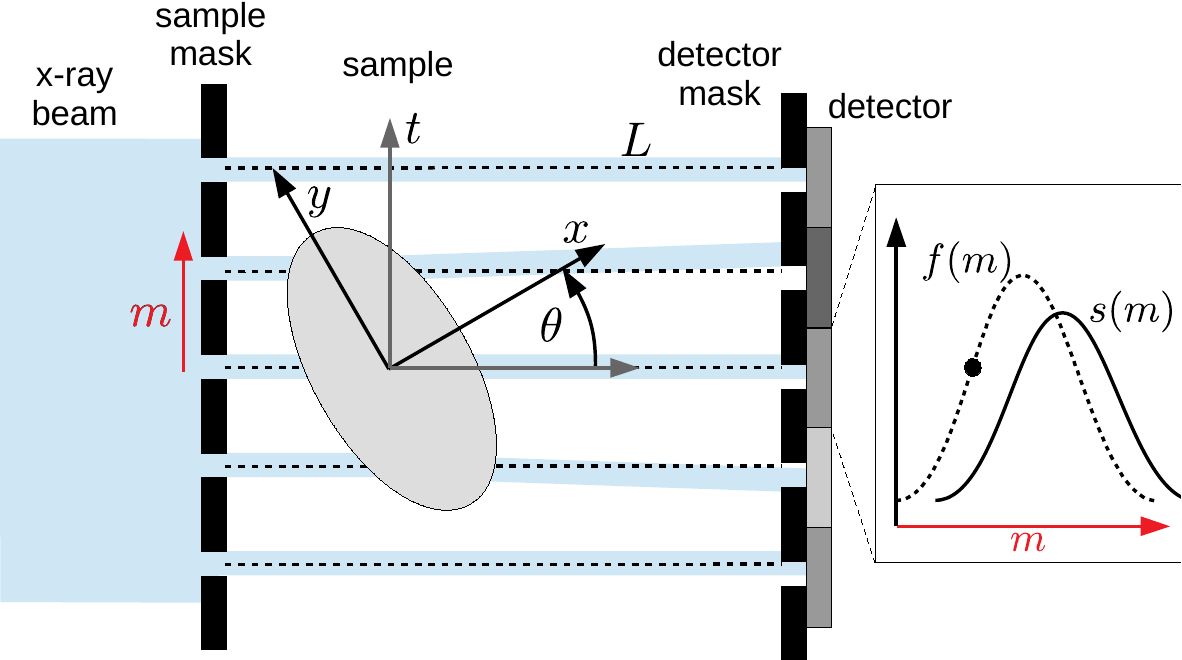}
\end{center}
\caption{Sketch of the edge-illumination setup and coordinate systems of interest. Inset: Typical illumination curves without sample $f(m)$ (dotted line) and with sample $s(m)$ (continuous line) obtainable by a lateral scan of the sample mask.}
\label{fig:setup}
\end{figure}

In EI, differential phase contrast is achieved by the utilization of two apertured masks (Fig.~\ref{fig:setup}). The sample mask confines the incident X-ray beam into smaller beamlets, which are refracted by the sample. This results in a lateral offset of the beamlets at the position of the detector, where the offset is transformed into a detectable intensity contrast by the detector mask, whose apertures cover a fraction of the pixels. Scanning the sample mask over one period (i.e., varying $m$) without a sample present in the beam provides the flat field illumination curve (IC), $f(m$). Repeating this scan with a sample present in the beam will yield the sample IC, $s(m)$, with a decreased intensity due to absorption and a shift due to refraction (i.e., derivative of the phase signal).

Recently, we have established a model-based iterative approach to tomographic reconstruction for EI~\cite{Modregger2019a} that includes a potential drift of optical elements during a scan as well as a heuristic parameter for ring artifact suppression. In the following, we will identify a cause of occurring ring artifacts, use this understanding to improve both experimental scan protocols as well as the speed of tomographic reconstruction. Further, we report on improvements in our numerical implementation, which in combination with the updated scan protocol leads to a significant improvement of reconstruction quality and speed.

\section*{Theoretical basis}

To begin with, we will collect the essential parts of the theoretical basis of our previously published algorithm for the reader's convenience. Tomographic reconstruction is based on the Radon transform~\cite{Kak1988}, which is given as the line integral
\begin{equation}\label{eq:radon}
\mathcal{R}\left[h(x,y)\right](t,\theta) = \int_{L} h(\mathbf{x})\, d|\mathbf{x}| 
\end{equation}
for an arbitrary, two-dimensional function $h(x,y)$ with the coordinate system $(x,y)$ fixed to the sample and $\mathbf{x}=(x,y)$. $L$ defines the lines of integration, which is parameterized by the pixel position at the detector $t$ and the projection angle $\theta$ (see Fig.~\ref{fig:setup}).

For EI, an IC without a sample present in the beam, $f(t,m)$, is acquired while varying $m$, the lateral offset of the sample mask position. Introducing the sample into the beam modifies the IC in two ways. First, the detected intensity is decreased by attenuation, which is given as the exponential of the negative projected linear attenuation coefficient $\mu(x,y)$: $\exp (-\mathcal{R}[\mu (x,y)])$, where the dependency of $\mathcal{R}$ on $(t,\theta)$ was omitted. Second, the IC is shifted according to the first derivative of the projected phase signal: $z\partial_t \mathcal{R}[\delta (x,y)]$ with $\delta(x,y)$, the refractive index decrement and $z$, the distance between sample and detector. Thus, the observable IC with sample, $s(t,\theta,m)$ is
\begin{equation}
s(t,\theta,m) = \exp (-\mathcal{R}[\mu (x,y)]) f(t,m-z\partial_t \mathcal{R}[\delta (x,y)]).
\end{equation}

It is well established that the combined reconstruction of absorption and phase data from single images in PPI can lead to a substantial improvement of contrasts in tomograms~\cite{Paganin2002a}. This approach is valid for homogeneous samples, which allows for assuming a linear relationship between the absorption and phase signal: $\mu (x,y) = \gamma^{-1} \delta(x,y) = h(x,y)$. Applying this assumption to EI provided the possibility to reconstruct $h(x,y)$ from on a single offset position $m$ by transforming the projections according to~\cite{Diemoz2015f,Diemoz2017}
\begin{equation}\label{eq:single_shot}
h(t,\theta) \propto \gamma^{-1} \log \left[ \mathcal F^{-1} \left[ \frac{ \mathcal{F}
\left[ s(t,\theta,m)/f(t,m)\right] }{1+i q z \gamma^{-1} f'(m)/ f(m) }
 \right]  \right]
\end{equation}
with $\mathcal{F}$ the Fourier transform with respect to $t$, its inverse $\mathcal{F}^{-1}$ and $q$ the variable conjugate to $t$. Here, $f(m)$ denotes the average of $f(t,m)$ over pixel positions $t$ and $f'(m)$ is the first derivative of $f(m)$.

Continuing with the theoretical basis for iterative reconstructions, we accounted for a potential drift of the sample mask position by introducing the projection angle dependent offset $m_o(\theta)$ as well as a heuristic additional degree of freedom for ring artifact suppression $m_r(t)$. The final forward model was given as
\begin{equation}\label{eq:modmodel}
s_{\mathrm{mod}}(t,\theta,m)  = \exp\left(-\mathcal{R}\left[h\right]\right)
\times f\left(t,m-m_o(\theta)-m_r(t)-z\gamma \partial_t \mathcal{R}\left[h\right]\right),
\end{equation}
where the dependency of $h$ on $(x,y)$ was suppressed for ease of readability. Retrieving the combined contrast $h(x,y)$ from the experimentally obtained sinograms $s_{\mathrm{exp}}(t,\theta,m)$ was then achieved by the iterative minimization of the cost function
\begin{equation}\label{eq:costfunction}
S = \sum_{t,\theta,m}||s_{\mathrm{exp}}(t,\theta,m)-s_{\mathrm{mod}}(t,\theta,m)||^2 + \lambda ||h(x,y)||^2,
\end{equation}
where the second term constitutes a regularization term that minimizes noise with an L2-norm. We used the limited memory Broyden--Fletcher--Goldfarb--Shannon (L-BFGS) implementation~\cite{Byrd1995a} in the SciPy library for Python~\cite{Jones2001} to minimize the cost function. The analytical gradients of $S$, which are essential for fast computation, as well as more details about the algorithm can be found in~\cite{Modregger2019a}.

At this point, we would already like to point out a potential ambiguity in the forward model (eq.~\ref{eq:modmodel}). If a solution to eq.~(\ref{eq:costfunction}) was found with some $m_o(\theta)$ and $m_r(t)$ then $m_o(\theta)+a$ and $m_r(t)-a$ with an arbitrary real number $a$ would also be a solution. This ambiguity tends to slow down the iteration process (demonstrated below) and should be avoided.

\section*{Experimental improvements}

Possible causes for the occurrence of ring artifacts in X-ray absorption tomography include defective detector pixels, damaged scintillator screens or beam instabilities~\cite{Munch2009a}. For EI, the measurement of the flat field IC, $f(t,m)$, constitutes an additional challenge. In the following, we will demonstrate that undersampling in terms of measured number of photons of the flat field IC leads to ring artifacts in the tomographic reconstructions.

The experiment was performed with the laboratory-based EI implementation at University College London~\cite{Hagen2014a}. We have imaged three different types of plastic granules (i.e., polystyrene (PS), polypropylene (PP) and poly methyl methacrylate (PMMA)). A Rigaku M007 rotating anode (Rigaku Corporation, Japan) with a Mo target and operated at 40~kV and 20~mA was used as the X-ray source. The detector was a Hamamatsu C9732DK flat panel (Hamamatsu, Japan) featuring a pixel size of 50~$\mu$m. Both masks were manufactured by electroplating of Au on a graphite substrate (Creatv Microtech Inc., USA) with a pitch of 38~$\mu$m for the sample mask and 48~$\mu$m for the detector mask, respectively. Total setup length was 85~cm and the sample to detector distance was 20~cm.

In order to obtain the experimental flat field ICs $f(t,m)$ the sample mask was scanned over one period with 33 steps. The scan was repeated 20 times to improve the sampling in terms of photon counts of $f(t,m)$ while taking mask positioning accuracy into account. The experimental sinograms $s_{\mathrm{exp}}(t,\theta,m)$ were measured with the sample mask on a slope position of the ICs (i.e., $m=9\,\mu$m offset from the maximum position; see thick black dot in Fig.~\ref{fig:setup}). Continuous rotation with a speed of 0.9~deg/s was employed over 360~degrees, which lead to a tomography scan time of 500~s. During the scan, 400 projections were acquired with an exposure time of 1.25~s each.

In the following, we will compare two data sets, one with 1 flat field scan and one with 20 repeated flat field scans, and three tomographic reconstructions approaches: single shot, iterative reconstruction with ring suppression ($m_r\not\equiv 0 $) and iterative reconstruction without ring suppression ($m_r\equiv 0 $). For the single shot approach data analysis was performed by applying eq.~(\ref{eq:single_shot}) with $\gamma=5$ to the raw data and subsequent tomographic reconstruction was done by using the inverse Radon transform provided in scikit-image~\cite{van2014a}. For the iterative approach (eq.~\ref{eq:costfunction}) the same $\gamma$ and $\lambda = 10^{-2}$ was used. The retrieved tomographic slice consisted of $N_t\times N_t=220\times 220$ voxels.

\begin{figure}[htbp]
\begin{center}
\begin{tabular}{rccc}
& single shot & iteration ($m_r\not\equiv 0$) & iteration ($m_r\equiv 0$) \\
\rotatebox{90}{\hspace*{10mm}1 flat scan} &
\includegraphics[width=0.29\textwidth]{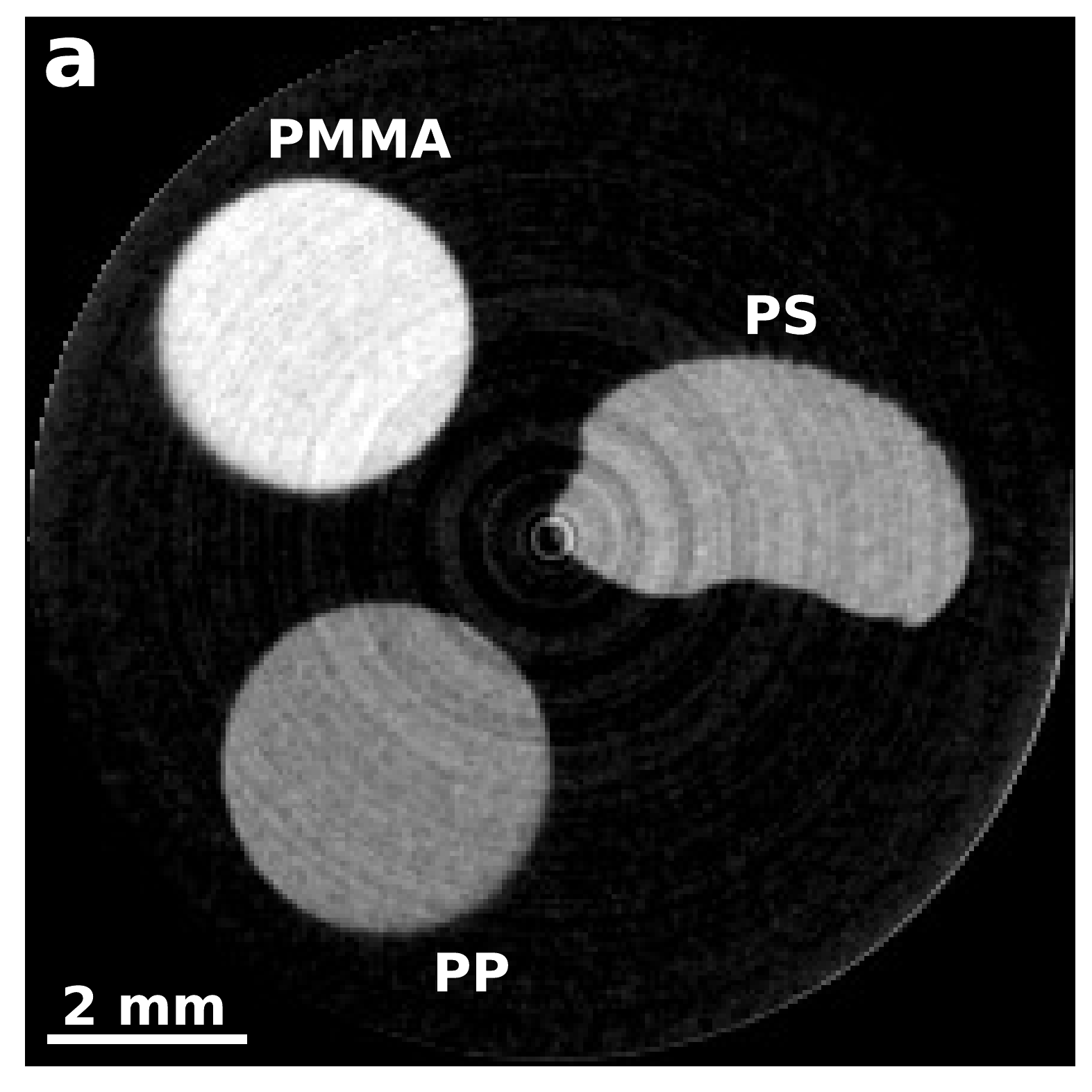} &
\includegraphics[width=0.29\textwidth]{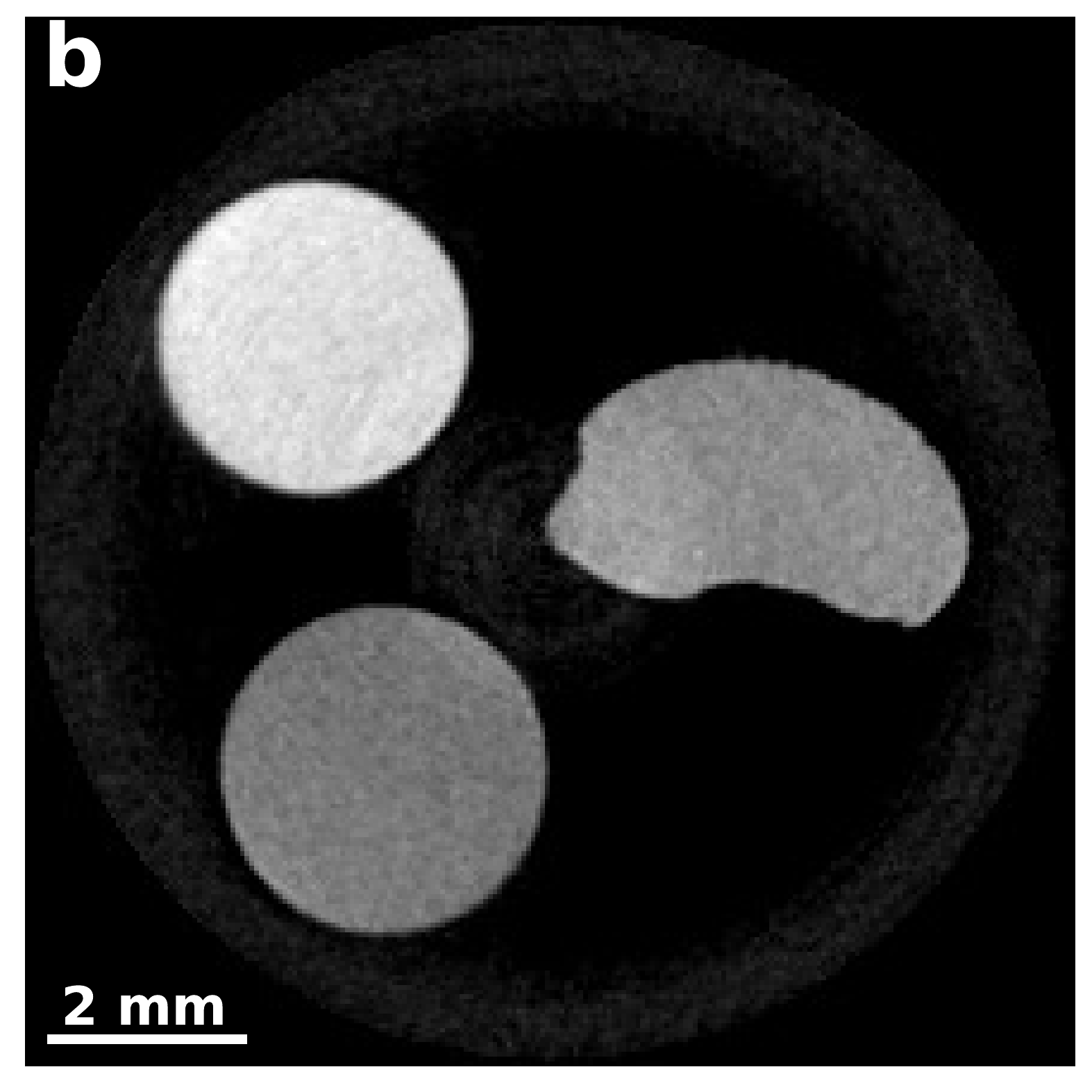} &
\includegraphics[width=0.29\textwidth]{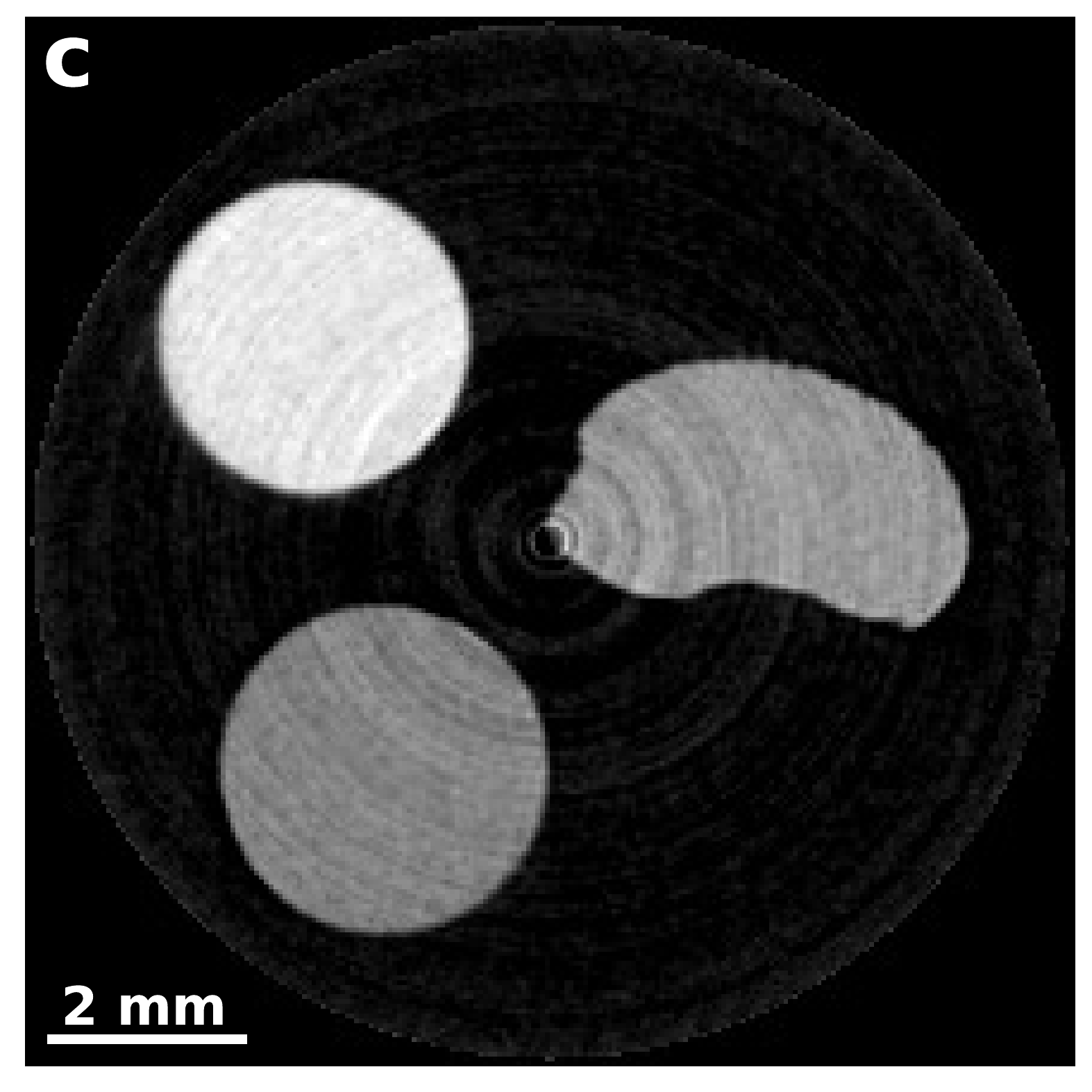} \\
\rotatebox{90}{\hspace*{10mm}20 flat scans} &
\includegraphics[width=0.29\textwidth]{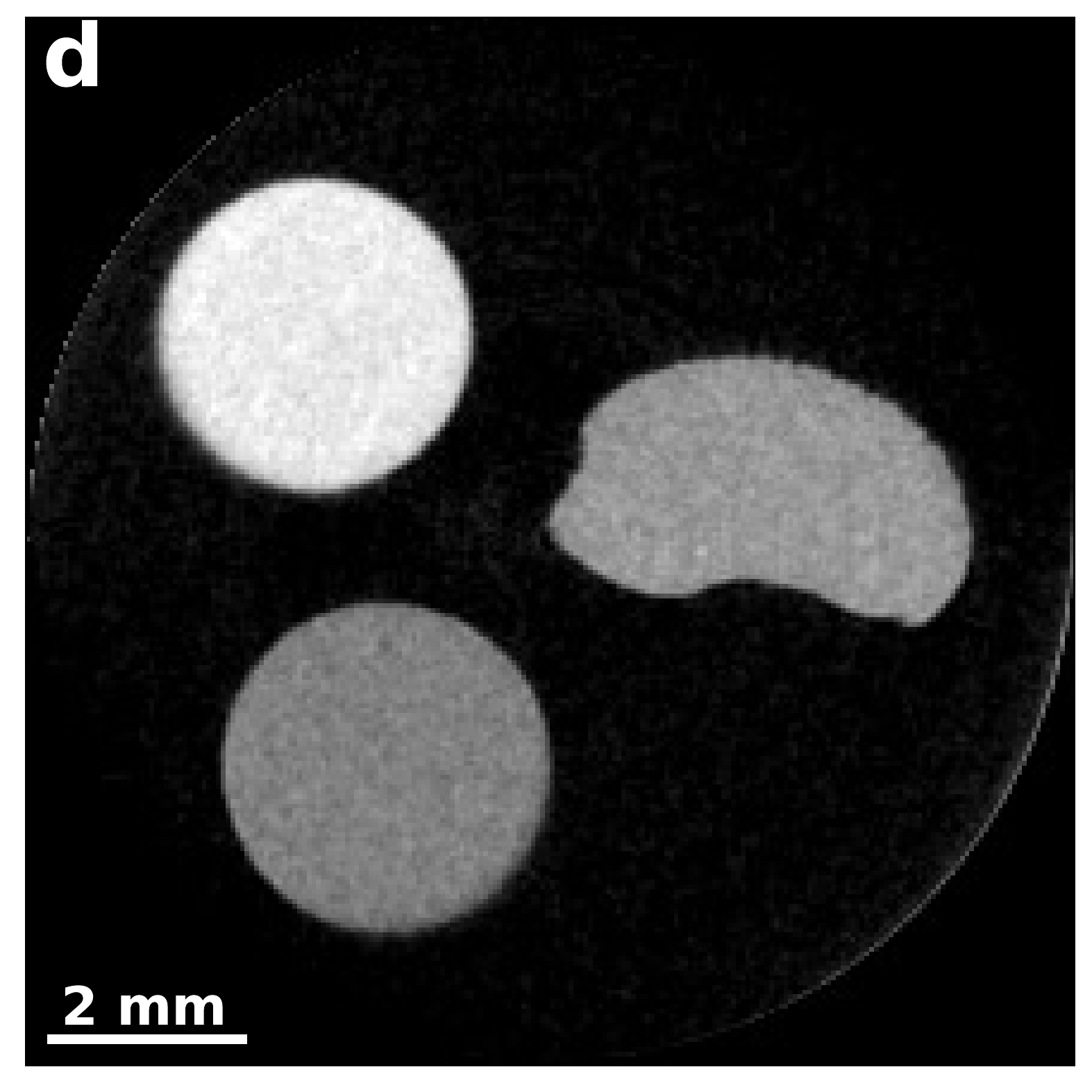} &
\includegraphics[width=0.29\textwidth]{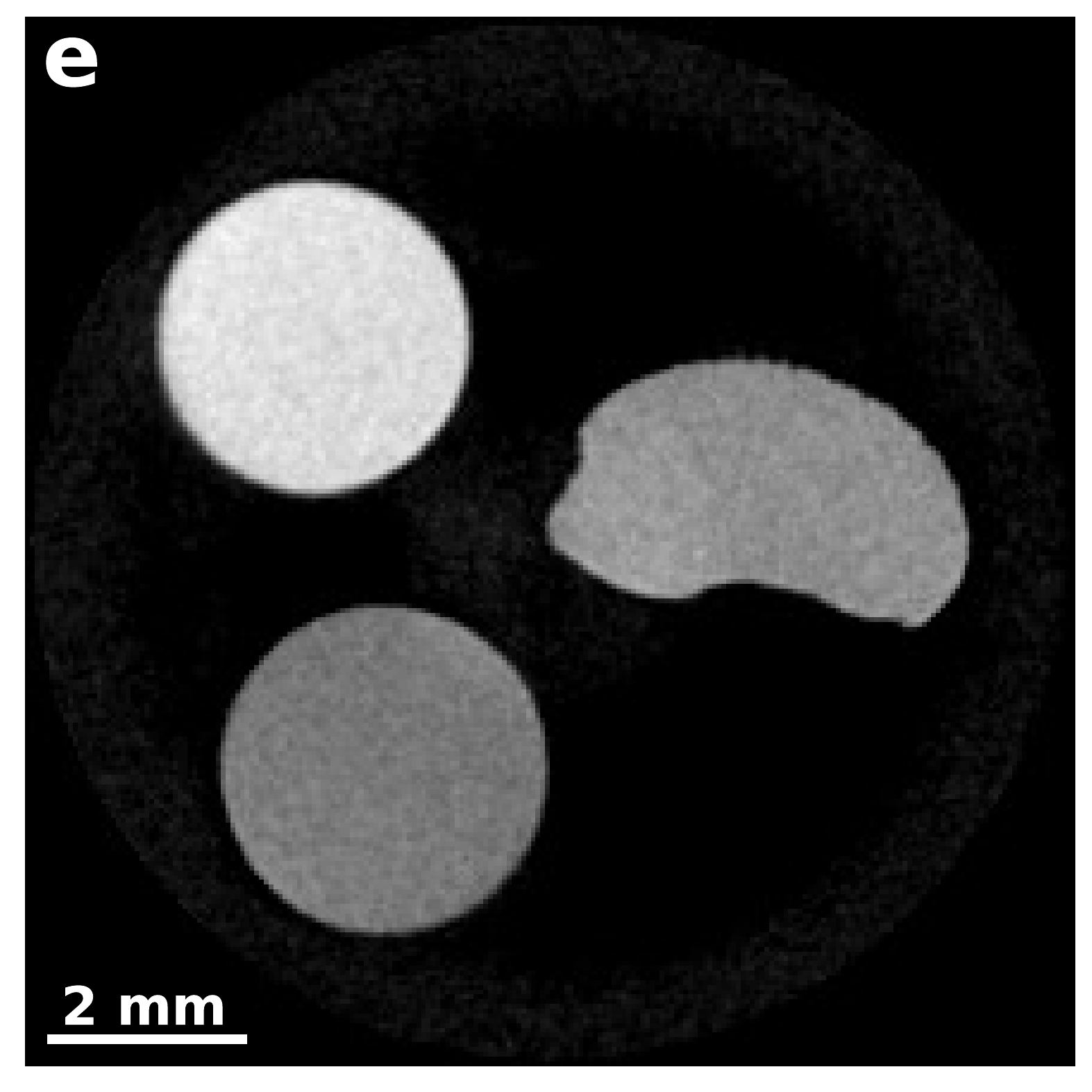} &
\includegraphics[width=0.29\textwidth]{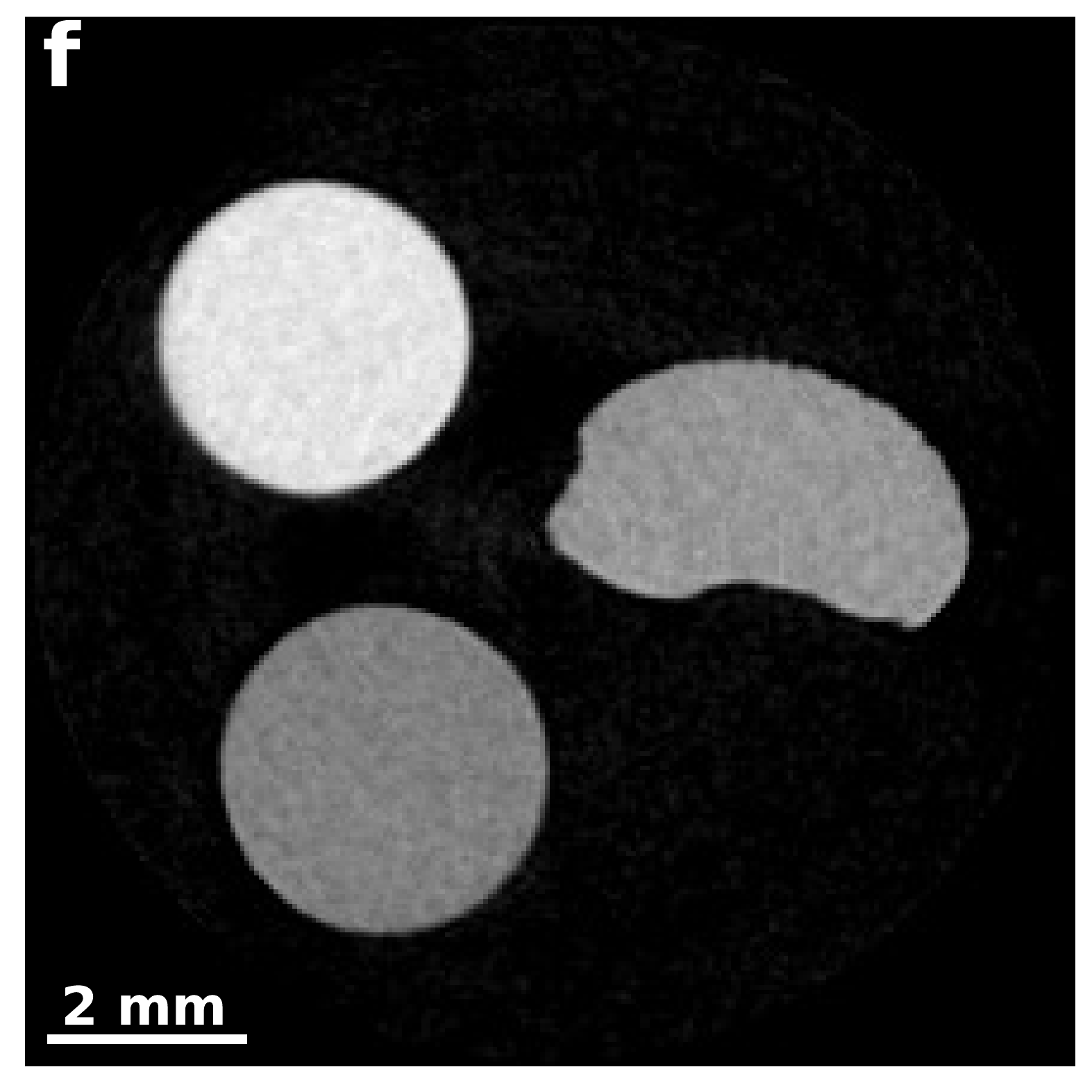} \\
\end{tabular}
\end{center}
\caption{Tomographic reconstruction of three different plastics granules using 1 flat scan (top row, a-c) or 20 flat scans (bottom row, d-f) with the single shot approach (left column, a \& d), iteration with ring suppression (middle column, b \& e) and iteration without ring suppression (right column, c \& f). All images share the same color map. The  large circular artifact in the single shot reconstructions is due to drift of the sample masks during the scan. It is evident that using repeated measurements of the flat field IC removes ring artifacts (cmp. a \& d), which implies that ring artifact suppression in the iterative forward model can be forgone (cmp. b \& f).}
\label{fig:recons}
\end{figure}

Figure~\ref{fig:recons} shows the results of the tomographic reconstructions using an undersampled flat field IC (top row; only 1 scan of the flat field IC was used) and an appropriately sampled flat field IC (bottom row; the average of 20 flat field scans were used). Comparing the results of the single shot approach (cmp. a to d) already makes it apparent that the cause of ring artifacts is the undersampling of the flat field ICs in terms of number of acquired photons. This can be explained as follows. Naturallym, the accuracy of the measured flat field ICs $f(t,m)$, which constitutes a necessary input for single shot reconstruction (eq.~\ref{eq:single_shot}), is limited by photon shot noise. Thus, in case of insufficiently measured number of photons, the accuracy of $f(t,m)$ will vary noticeably with the detector pixel $t$. Since these inaccuracies are identical for all projection angles $\theta$, they will lead to stripes in the sinograms and, consequently, ring artifacts in the tomographic reconstructions.

For the presented iterative approach the utilization of an adequately sampled flat field IC opens the possibility to remove the ring artifact suppression from the forward model (i.e., $m_r(t)\equiv 0$ in eq.~\ref{eq:modmodel}). The comparison of Fig.~\ref{fig:recons}b (i.e., iterative reconstruction using 1 flat field scan and ring artifact suppression, $m_r(t)\not\equiv 0$) and Fig.~\ref{fig:recons}f (i.e., iterative reconstruction using 20 flat field scans and without ring artifact suppression, $m_r(t)\equiv 0$) shows that this can be done while preserving the evident quality of the reconstructions (see also quantitative comparison in Tab.~\ref{tab:cnr}). Simultaneously, convergence of iterative reconstruction without ring suppression is achieved in only 36 iteration steps compared to 87 iteration steps with ring suppression in this example. Since the computations related to $m_r(t)$ can be forgone and the above mentioned ambiguity corresponding to $m_o(\theta)$ is avoided, using an appropriate sampled flat field IC resulted in an increase in reconstruction speed by about the factor of 3. Naturally, total scan time (i.e., flat field plus sample IC) will be increased but the delivered dose will be identical.

Furthermore, the single shot retrieval, as given in eq.~(\ref{eq:single_shot}), implies filtering, which inevitably leads to image blurring and a worsening of spatial resolution. The iterative model (eq.~\ref{eq:modmodel}) on the other hand, lacks explicit filtering and image blurring only occurs on a neighbouring pixel-basis due to the numerical implementation of the derivative as finite differences. A visual comparison of Fig.~\ref{fig:recons}d and~f confirms that the iterative approach provided improved spatial resolution.

We have used Fourier ring correlation (FRC) in order to quantitatively compare the spatial resolutions~\cite{Koho2019}. FRC is based on the normalized Fourier transform of the cross-correlation between two images of the same region but with independent noise realizations. The corresponding histogram, $\mathrm{FRC(\omega)}$, is calculated by
\begin{equation}\label{eq:frc}
\mathrm{FRC(\omega)} = \frac{F_1(\omega)F_2(\omega)^*}{\sqrt{F_1(\omega)^2 F_2(\omega)^2}}
\end{equation}
with the spatial frequency $\omega$ and the Fourier transform of the individual images $F_1$ and $F_2$, respectively.

\begin{figure}[htbp]
\begin{center}
\includegraphics[width=0.6\textwidth]{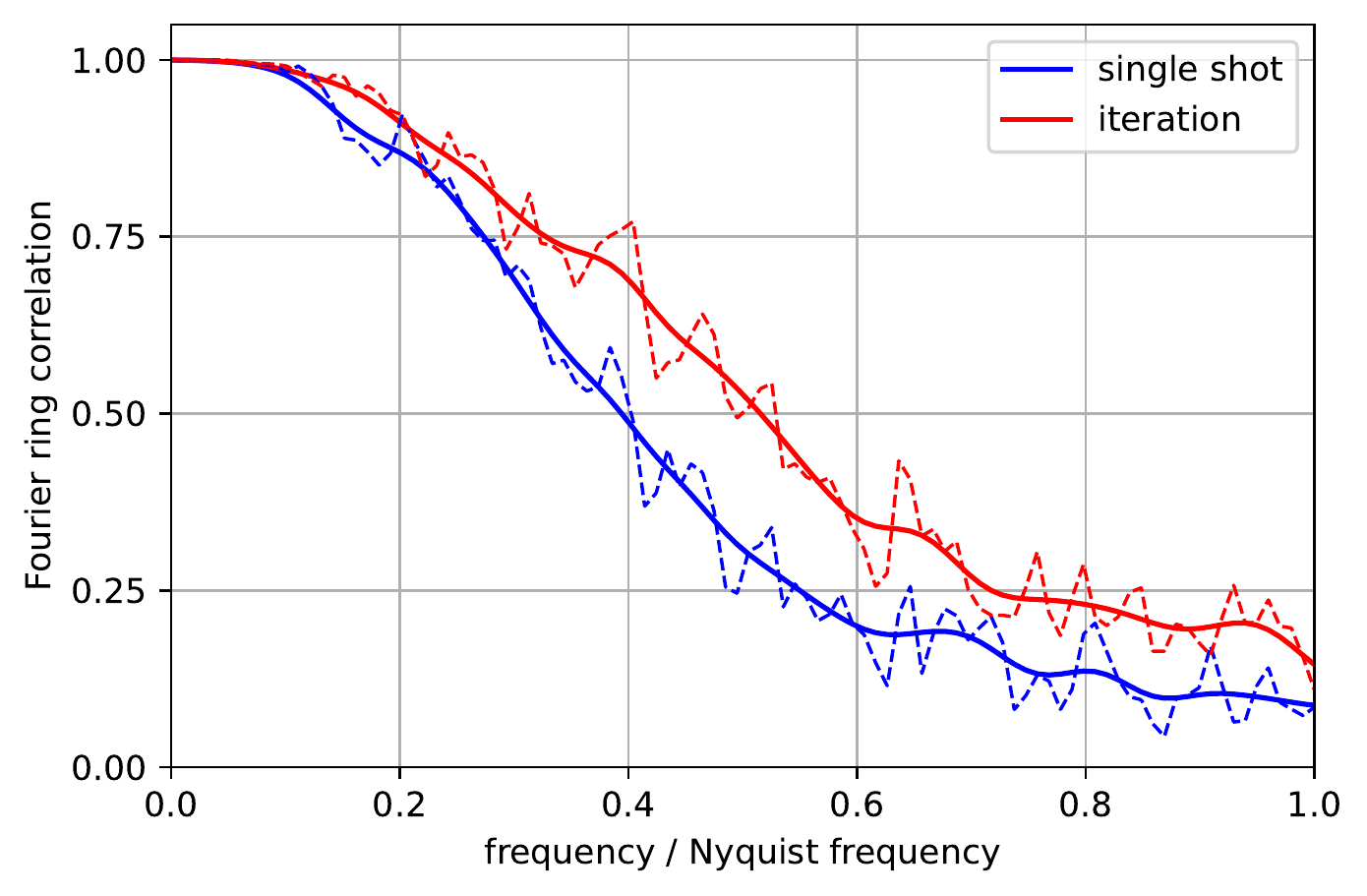} 
\end{center}
\caption{Comparison of the Fourier ring correlation for the tomographic reconstruction based on the single shot and on the presented iterative approach. The spatial resolution of the iterative approach was superior by 30\%.}
\label{fig:frc}
\end{figure}

To calculate the FRC, we have split the sinograms into odd and even projections (200 projections each) and used single shot as well as iterative tomographic reconstruction with an identical proportionality factor $\gamma = 5$ in order to provide independent noise realizations. As input for FRC we have cut a square area of 140 by 140 pixels out of the center of the resulting reconstructions to avoid the influence of the circular artifact visible in Fig.~\ref{fig:recons}d. The resulting histograms have been smoothed by a Gaussian and are shown in Fig.~\ref{fig:frc}. Using a conservative cutoff value of 0.5, we found that tomographic reconstruction based on single shot retrieval provided a spatial resolution of $245\,\mu$m (4.9~pixels) while iterative reconstruction achieved $190\,\mu$m (3.8~pixels). This  improvement of 30\% for the presented iterative approach proved insensitive towards variations in $\gamma$, cutoff value and iteration parameters such as number of iteration steps or the regularization parameter $\lambda$.



\section*{Algorithmic improvements}

With this publication, we are making the source code of this iterative approach to tomographic reconstruction of X-ray imaging with edge-illumination available to the public~\cite{Modregger2022a}. Thus, we would like to report on several significant improvements in the numerical implementation compared to our previous publication~\cite{Modregger2019a}.

\begin{table}[htbp]
\begin{center}
\begin{tabular}{ll|lll}
$N_t$ & $N_{\theta}$ & lookup & Numba & OpenMP \\ \hline\hline
90 & 45 & 0.74 & 0.18 & 0.27 \\
180 & 90 & 3.5 & 1.4 & 1.2 \\
360 & 180 & 18 & 11 & 5.6\\
720 & 360 & 100 & 88 & 27\\ 
\end{tabular}
\end{center}
\caption{Relative computation times of three different algorithms for a single Radon transform. Lookup refers to our previously published algorithm that utilized a lookup table, Numba and OpenMP, where the latter used a modest 3 threads, refer to the proposed algorithms. The size of the calculated sinograms are given by $N_t$, the number of detector pixels and $N_{\theta}$, the number of projection angles. The performance of the lookup algorithm for $N_t = 720$ detector pixels was set to 100.}
\label{tab:recon}
\end{table}

During iteration the calculation of the Radon transform~(eq.~\ref{eq:radon}) constitutes the most time consuming part, which takes up about 85\% of the computation time. In this update, we eliminated several superfluous calls of the Radon transform within each iteration step, which resulted in a equivalent speed up. Further, our previously published approach utilized a lookup table to compute the Radon transform, which has a high demand for computer memory. For example, in our implementation the lookup table for $N_t = 360$ detector pixels and $N_{\theta} = 180$ projection angles took up around 2~GB of memory.  In order to reduce computer memory requirements and to improve computational speed, we implemented the Radon transform in both Python Numba with a just-in-time compiler that generates fast machine code~\cite{Lam2015} as well as in C-based OpenMP~\cite{Dagum1998}. Instead of a lookup table the indices for the linear interpolation used to compute the Radon transform are calculated on the fly, which renders memory usage negligible. Further, the comparison of computation times for a single Radon transform of the three algorithms given in Tab.~\ref{tab:recon} shows a modest to a very significant speed up depending on the size of the sinograms.


\section*{Combined results}

We assessed the performance advancement due to the combination of experimental and algorithmic improvements on the iterative reconstructions in terms of computational speed, number of iteration steps until convergence and the contrast to noise ratio (CNR) between the materials at hand. We used the following definition of the CNR
\begin{equation}\label{eq:cnr}
\mathrm{CNR} = \frac{|\bar{h_1}-\bar{h_2}|}{\sqrt{\mathrm{std}(h_1)^2+\mathrm{std}(h_2)^2}},
\end{equation}
where $\bar h_i$ denotes the mean of area $i$ and $\mathrm{std}(h_i)$ the corresponding standard deviation. Here, we determined the CNR between the different materials by circular areas with a radius of 40 voxels located in the center of the plastic granules.

\begin{table}[htbp]
\begin{center}
\begin{tabular}{r|r|rrrr}
\makecell{number of\\flat scans} & quantity & \makecell{lookup \\ ($m_r\not\equiv 0$)} &  \makecell{lookup\\ ($m_r\equiv 0$) } & \makecell{Numba\\($m_r\not\equiv 0$)} &  \makecell{Numba\\ ($m_r\equiv 0$)}\\ \hline\hline
1 & rel. iteration time & {\bf 100}  & 36.5 & 17.4 & 7.31\\
  & iteration steps & {\bf 112} & 40 & 87 & 37\\
  & CNR PMMA-PP   & {\bf 10.1}   & 9.5   & 10.7   & 9.5\\ 
  & CNR PMMA-PS   &  {\bf 7.0}   & 6.6   & 7.3    & 6.6\\
  & CNR PS-PP     &  {\bf 2.2}   & 1.8   & 2.3    & 1.8\\ \hline   
20& rel. iteration time & 31.1  &  58.4 & 12.3 & {\bf 6.62}\\
  & iteration steps & 36 & 68 & 51 & {\bf 36}\\
  & CNR PMMA-PP  & 12.2   & 12.6  & 12.6   & {\bf 12.6} \\
  & CNR PMMA-PS  & 8.5    & 9.6   & 9.3    & {\bf 9.6} \\
  & CNR PS-PP    & 2.6    & 2.6   & 2.8    & {\bf 2.6}\\
\end{tabular}
\end{center}
\caption{Performance comparison between the previously published approach and the combination of experimental and algorithmic improvements presented here. Lookup and Numba refer to the implementation of the Radon transform used in the iteration algorithm. $m_r \not\equiv 0$ and $m_r \equiv 0$ refer to the usage and lack thereof of ring artifact suppression in the forward model (eq.~\ref{eq:modmodel}). Computations times are given relative to the entry with a value of 100. Comparing the entries of particular interest, printed in bold, demonstrates that the combined improvements enabled a 15-fold computational speed up of the iteration while improving the CNR by at least 20\%.}
\label{tab:cnr}
\end{table}

Table~\ref{tab:cnr} benchmarks the performance of the combined experimental and algorithmic improvements described. The results of single shot reconstruction are not shown as it was already demonstrated in a previous publication that the iterative approach provides higher CNR values~\cite{Modregger2019a}. Computations times are given as relative to the entry with a value of 100, which was 219~s on a standard work station with an i5 CPU running at 2.10~GHz. The results clearly demonstrate that the combined improvements enabled a 15-fold computational speed up of the iteration while modestly improving the CNR by 20\% to 30\%. Using the OpenMP instead of the Numba implementation of the Radon transform would only result in a modest additional increase in speed for the number of detector pixels at hand (see Tab.~\ref{tab:recon}).

In order to provide an indication of total iteration times for different reconstruction sizes, we have used numerical simulations with the same parameters as described above. We found that the total iteration time was 31~s for $N_t=N_{\theta}=360$, 290~s for $N_t=N_{\theta}=720$ and 2700~s for $N_t=N_{\theta}=1440$, respectively.


\begin{figure}[htbp]
\begin{center}
\includegraphics[width=0.147\textwidth]{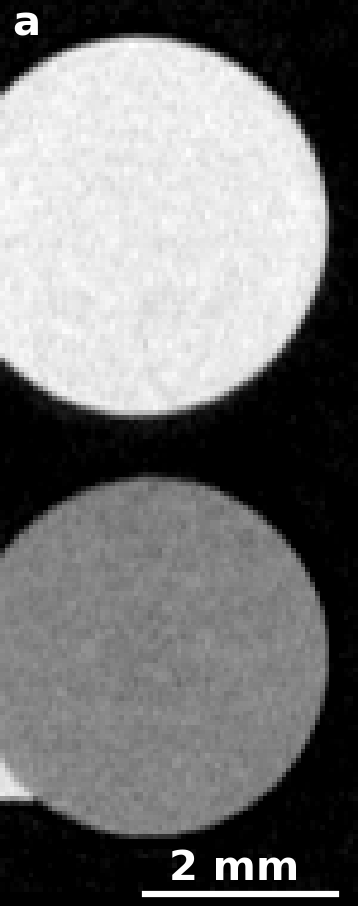}
\hspace*{2mm}
\includegraphics[width=0.6\textwidth]{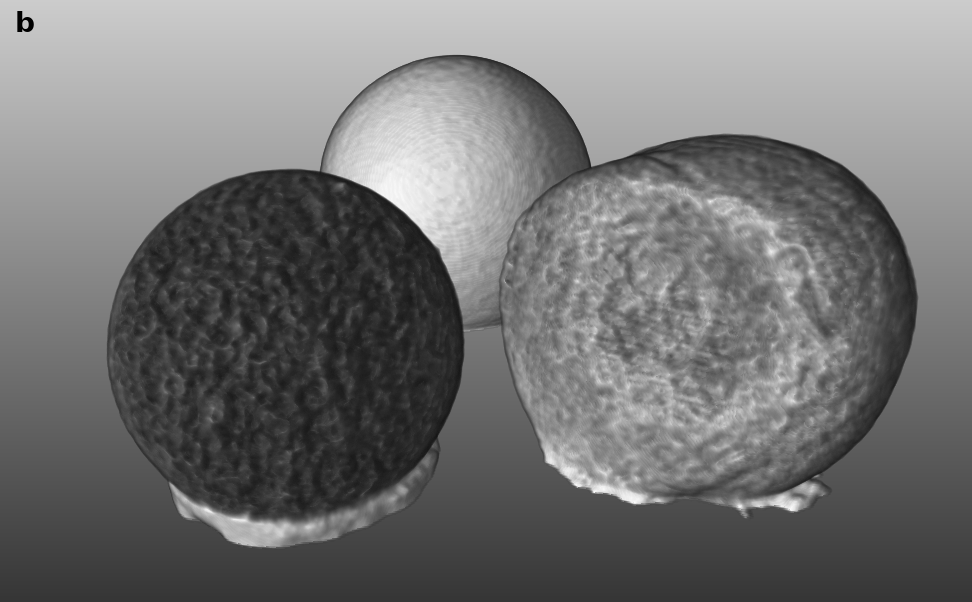}
\end{center}
\caption{Visualization of full three dimensional iterative reconstructions of the plastic granules. (a) A vertical cut through the data set demonstrating that the iterative approach performs reliably over different slices. (b) A three dimensional rendering of the full data set.}
\label{fig:3d}
\end{figure}

Finally, we applied the iterative approach to the full three dimensional stack of the plastic granules. The full reconstruction of the 220x220x75 voxels took 23~min in a single thread on the work station described above. As the presented iterative approach works independently on each 220x220 slice, there is the potential for strong slice by slice background variations~\cite{Massimi2021}, which could be hidden by presenting only single slices. Figure~\ref{fig:3d}a shows a vertical cut through the stack, where there is no noticeable variation in the background visible, which shows that the iterative algorithm works reliably with respect to stacking of individual slices. Figure~\ref{fig:3d}b displays a three dimensional rendering of the plastic granules, which was produced using MeVisLab~\cite{Ritter2011}.

\section*{Conclusions}

In conclusion, we have identified that in tomographic phase sensitive X-ray imaging based on the edge-illumination principle undersampled flat field ICs can lead to ring artifacts in the reconstructions. We have demonstrated that ring artifacts all but disappear if adequately sampled flat field ICs are utilized (here: 20 repeated scans). This has opened the possibility to forgo the ring artifact suppression in the forward model of the presented iterative tomographic reconstruction. Consequently, the iteration was about 3 times faster while preserving image quality. In addition, we have demonstrated that the iterative approach provides improved spatial resolution when compared to non-iterative, single shot-based reconstruction.

Further, we have reported on several significant improvements to our algorithmic implementation, which is now available online~\cite{Modregger2022a}. We have demonstrated that the com\-bination of experimental and algorithmic improvements lead to a 15-fold speed up and a CNR improvement of 20\% for iterative reconstruction in the case at hand. Finally, we have shown that the reliability of iterative reconstruction by the absence of slice to slice variations in three dimensional stacks.

\section*{Funding}
AO was supported by the Royal Academy of Engineering under the Chairs in Emerging Technologies scheme. Additional funding was provided by Perkin Elmer Inc and by EPSRC (Grant EP/T005408/1). ME was supported by the Royal Academy of Engineering under the RA Eng Research Fellowship scheme.

\section*{References}

\end{document}